\documentclass[aps,twocolumn,amsmath,amssymb,showpacs,showkeys,floatfix]{revtex4}
\usepackage{graphicx}
\usepackage{amssymb} 

\begin{document}




\title{Hierarchy in Mid-Rapidity Fragmentation: Mass, Isospin, Velocity Correlations}

\author{V.Baran$^{1,2,*}$, M.Colonna$^{3}$,  M. Di Toro$^{3,4}$, R. Zus$^{1}$}

\affiliation{$^{1}$ Physics Faculty, University of Bucharest, Romania\\
$^{2}$ NIPNE-HH, Bucharest-Magurele, Romania \\ 
$^{3}$ Laboratori Nazionali del Sud INFN, I-95123 Catania, Italy \\ 
$^{4}$ Physics and Astronomy Dept., University of Catania\\
$^{*}$ email: baran@ifin.nipne.ro}

\begin{abstract}
We present new features of the transition from nuclear multifragmentation to
neck fragmentation in semi-central heavy-ion collisions at Fermi energies
as obtained within a microscopic transport model, Stochastic Mean Field (SMF).
We show that along this transition specific hierarchy phenomena
of some kinematic observables associated with the intermediate mass fragments develop.
Their correlations with the dynamics of isospin degree of freedom, predicted by our calculations, 
open new possibilities to learn about the density dependence of nuclear symmetry energy
below saturation, as well as about the fragmentation mechanisms. Detailed results are
presented for mass symmetric $Sn+Sn$ reactions with different isospin content at $50A$ $MeV$.
\end{abstract}


\pacs{25.70.Pq, 25.70.Mn, 21.65.Ef, 24.10.Cn}

\keywords{
Nuclear fragmentation, Symmetry energy, Reaction mechanisms 
}

\maketitle

\section{Introduction}

 Nucleus-nucleus collisions provide a unique tool to explore the properties of
finite interacting fermionic systems in a broad range of densities and temperatures.
At energies between $10$ and $100A$ $MeV$, usually referred to as 'Fermi energies',
the mean-field and collisional effects are quite balanced leading to a very intricate
dynamics, sensitive to impact parameter and beam energy. 
Entrance channel effects, as well as phenomena well explained in terms of statistical 
equilibrium, can coexist. Moreover, as a consequence of the two-component character of 
nuclear matter, additionally features due to isospin manifest. Indeed, the symmetry 
energy term in the equation of state (EOS) was one of the main subjects of interest 
during the last decade \cite{bar05a,ste05,bao08}.

  The fragmentation process is an ubiquitous phenomenon observed at Fermi energies.
However, the underlying reaction mechanisms can be rather different and a detailed study 
can provide independent information on the nuclear EOS out of saturation. The aim of this paper
is to suggest new fragment mass-velocity-isospin correlations particularly sensitive to
the various mechanisms, as well as to the in-medium nuclear interaction.

  For central collisions, the nuclear multifragmentation can be associated with
a liquid-gas phase transition in a composite system \cite{colonna2004}. 
While the final state configurations are well described within statistical
equilibrium models \cite{Rad05}, but also within hybrid models coupling a dynamical
formation and evolution of primary fragments with a secondary decay stage
 \cite{colonnabob}, the kinetics of this phase transition can be related to spinodal
decomposition in two-component nuclear matter \cite{colonna2004,bar98,baran2001}
accompanied by the isospin distillation. Increasing the impact parameter, the neck
fragmentation, with a peculiar intermediate mass fragments (IMF, $3 \le Z \le 20$ )
distribution and an entrance channel memory, was observed experimentally
\cite{defilippo05a,ditoro06, milazzo02,russotto10} and predicted by various
transport models \cite{bar04},\cite{papa07}. In this case the low-density neck
region triggers an isospin migration from the higher density regions corresponding
to projectile-like fragment, PLF, and target-like fragment, TLF. 
Therefore, the isospin content of the IMF's is expected to reflect
the isospin enrichment of mid-velocity region. For even more peripheral collisions,
an essentially binary reaction in exit channel can by accompanied by a dynamically induced
fission of the participants \cite{defilippo05b,russot06}
and for $N/Z$ asymmetric entrance channel combinations isospin diffusion takes place
\cite{tsang2004,bar05,rizzo2008}.

 Consequently, the isospin degree of freedom can be seen as a precious tracer providing
additional information about the physical processes taking place during the
evolution of the colliding systems. Moreover, from a comparison between the
experimental data and the theoretical model predictions, isospin dynamics allows 
to investigate the density and/or temperature dependence of the symmetry energy. 
More exclusive analysis from the new experimental facilities certainly will 
impose severe restrictions on various models and parametrization 
concerning this quantity.

Following these arguments, the purpose of this article is to inquire on the dynamics of 
the fragmentation process from semi-central to semi-peripheral collisions. 
We explore the  kinematic properties of the fragments produced at the
transition from multifragmentation to neck fragmentation. A hierarchy in the
transverse velocity of IMF's is clearly evidenced. Moreover, new interesting 
correlations between kinematic features of the fragments and isospin dynamics, 
which can provide clues in searching for the most sensitive observables to
the symmetry energy, are noted. We mention that for central collisions a radial
 expanding multifragmenting source develop.
In this case a correlation between the $N/Z$ of the fragments
and their kinetic energy, sensitive to the density behavior of the 
symmetry energy, was recently evidenced in a transport model \cite{colonna2008}.
The average value of this ratio decreases with the kinetic energy per nucleon 
and it is asy-EOS dependent.

 An experimental study of internal correlations for the fragmentation of quasi-projectiles
was performed by Colin et al. \cite{colin2003}, within the INDRA collaboration. 
For certain classes of events a hierarchy  of mass fragments along the beam axis
was interpreted in terms of the breakup of the very elongated structure emerging from the 
interaction of the two colliding nuclei. More recently, McIntosh and al. 
examined the fragment emission from $Xe+Sn$ peripheral and mid-peripheral dissipative 
collisions \cite{mcintosh2010}. A significant enhancement 
of backward fragments yield relative to the forward
component as well as an alignment with the direction of projectile-like residue velocity
were evidenced.

In section II we briefly review the transport approach 
and specify the reactions which are studied. Section III is focused 
on the properties of the observed fragmentation mechanisms. Isospin effects are 
analyzed in section IV in connection with the kinematic features of the fragments.
Finally, in section V the conclusions and some suggestions for
experiments are presented.

\section{The transport approach}

To achieve the goal of this work we employ a semi-classical microscopic transport model, 
Stochastic Mean Field (SMF),
based on Boltzmann-Nordheim-Vlasov (BNV) equation \cite{baran2002}.
Our choice is motivated by the requirement
to have a well implemented nuclear mean-field dynamics
together with the effects of fluctuations induced by two-body scattering.
Experimental indications
at energies between 20 AMeV and 100 AMeV, including
the behavior of collective flows, suggest that mean-field plays an essential role
in shaping the evolution of the system.
Within the Stochastic Mean-Field model, the time evolution of the one-body
distribution function $f({\bf r},{\bf p},t)$ is described by a
Boltzmann-Langevin equation \cite{rizzo2008a}: 
\begin{equation}
\frac{\partial f}{\partial t}+\frac{\bf p}{m}\frac{\partial f}{\partial {\bf r}}-
\frac{\partial U}{\partial {\bf r}}\frac{\partial f}{\partial {\bf p}}=I_{coll}[f]+\delta I[f],
\end{equation}
where the fluctuating term $\delta I[f]$ is implemented through stochastic spatial 
density fluctuations \cite{colonna98a}.
The collision integral $I_{coll}[f]$ for fermionic systems takes into account the 
energy, the angular and isospin dependence of free nucleon-nucleon cross sections.

The symmetry energy effects were studied by employing two different density
parametrizations \cite{colonna98} of the  mean field:
\begin{eqnarray}
U_{q}&=&A\frac{\rho}{\rho_0}+B(\frac{\rho}{\rho_0})^{\alpha+1}
+C(\rho)
\frac{\rho_n-\rho_p}{\rho_0}\tau_q+ \nonumber  \\
&+&\frac{1}{2} \frac{\partial C}{\partial \rho} \frac{(\rho_n-\rho_p)^2}{\rho_0},
\end{eqnarray}
where $q=n,p$ and $\tau_n=1, \tau_p=-1$.
For asysoft EOS, $\displaystyle\frac{C(\rho)}{\rho_0}=482-1638 \rho$, 
the symmetry energy $E_{sym}^{pot}=\displaystyle\frac{1}{2}C(\rho) \frac{\rho}{\rho_0}$ 
has a weak density dependence close to the saturation, being 
almost flat around $\rho_0$. For asysuperstiff case,
$\displaystyle\frac{C(\rho)}{\rho_0}=\frac{32}{\rho_0}\frac{2 \rho}{\rho+\rho_0}$, 
the symmetry energy is quickly decreasing for densities below normal density.
The coefficients $A,B$ and the exponent $\alpha$, characterizing the isoscalar part of
the mean-field, are fixed requiring that the saturation properties of symmetric 
nuclear matter with a compressibility around $215$$ MeV$ are reproduced.

A comparative study of the reactions $^{132}Sn+^{132}Sn$ (EE system),
$^{124}Sn+^{124}Sn$ (HH) and $^{112}Sn+^{112}Sn$ (LL) at $50MeV/A$ is performed. 
The last two combinations were intensively analyzed in the recent years at MSU \cite{tsang2004}.
We shall focus at the value of impact parameter $b=4$ $fm$ for which a typical 
behavior corresponding to the transition from multifragmentation to neck fragmentation, 
a process not  very much investigated up to now, is clearly noted in our simulations. 
Indeed our previous results for these systems, indicate at $b=4 fm$ a memory of entrance channel, 
through the existence of well defined PLF's and TLF's,
even if the multiplicity of intermediate mass fragments is still quite large \cite{baran2002}.
At $b=6 fm$ the reaction mechanism corresponds to a neck
fragmentation with mostly two or three IMF's observed in the mid-rapidity region and a short
nucleus-nucleus interaction time. 

Therefore, along this transition region, for impact parameters between $3fm$ and $5fm$,
a mixing of features associated to multifragmentation and neck fragmentation are expected.
The relative values of interaction time (of the order of $120-140$ $fm/c$), of the
time associated to fragment formation and growth, as well as
of the time scales for isospin migration and 
distillation, will determine the properties of emitted IMF's. Consequently, a good
sensitivity to the symmetry energy density dependence can be expected.

\section{Fragmentation Mechanism}

A total number of 2000 events is generated for each 
entrance channel combination and equation of state at impact parameter $b=4 fm$.

\begin{figure}
\begin{center}
\includegraphics*[scale=0.325]{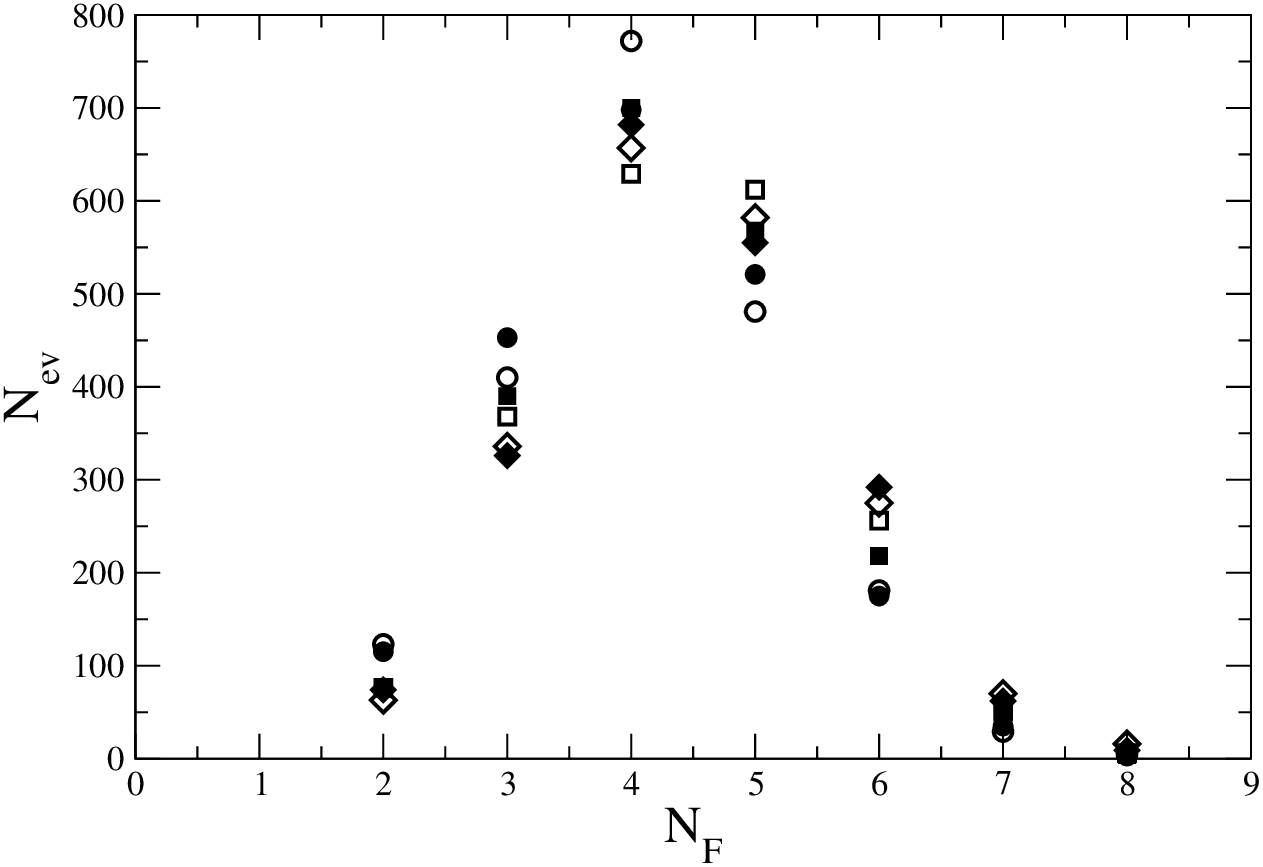}
\end{center}
\caption{Fragment multiplicity distribution at $b=4fm$. Circles:$^{112}Sn+^{112}Sn$.
Squares:$^{124}Sn+^{124}Sn$. Diamonds: $^{132}Sn+^{132}Sn$.
The filled symbols:asysoft EOS. The open symbols:asysuperstiff EOS.}
\label{mult}
\end{figure}

First, we adopt an analysis method of kinematic properties previously 
employed in studies concerning dynamical fission or neck fragmentation mechanisms 
\cite{stef1995,wilcz2005}. 

\begin{figure}
\begin{center}
\includegraphics*[scale=0.28]{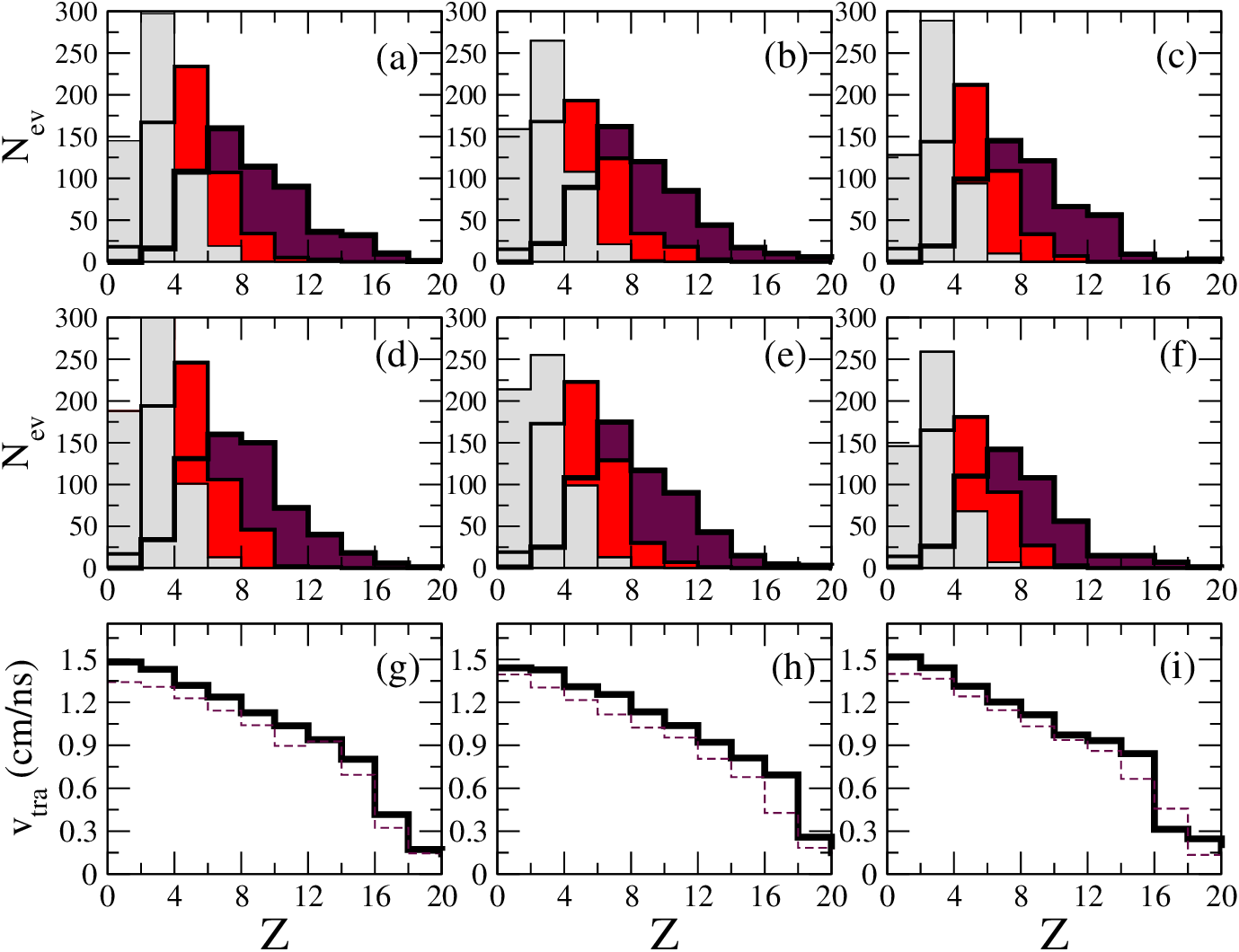}
\end{center}
\caption{The charge distribution for each IMF in hierarchy: asysoft EOS (upper row) and
asysuperstiff EOS (middle row). HH combination:(a),(d),(g). EE combination:(b),(e),(h).
LL combination:(c),(f),(i). 
Average transverse velocity distribution as a function
of the charge (bottom row) for asysoft EOS (thick-solid line) and asysuperstiff EOS (thin-dashed line).
All results refer to events with IMF multiplicity three. The histograms brighten as the rank 
of IMF's increases.
}
\label{dis5a}
\end{figure}

After the freeze-out time, corresponding to the saturation of the number of formed IMF's,
we propagate the Coulomb trajectories of all fragments until a configuration
where the Coulomb interaction becomes negligible. The asymptotic velocities of PLF and TLF 
define an intrinsic axis of the event by the vector 
${\bf{V}_r}= {\bf{V}}(H_1)-{\bf{V}}(H_2)$ always oriented from the second heaviest
fragment $H_2$ toward the heaviest one $H_1$. Even for mass symmetric  
entrance channels this is an appropriate definition when 
searching for the correlations between kinematic properties of the IMF's and the break-up
of the initial composite system. At freeze-out time the IMF's of each event 
are ordered in mass. The orthogonal and parallel components
of their asymptotic velocities with respect to the intrinsic axis, $v_{tra}$ and $v_{par}$,
together with their charge $Z$, are determined.
The events are classified according to the number of observed IMF's
at freeze-out time. We report in Fig.\ref{mult} the fragments multiplicity distributions 
associated to all studied cases. It is observed that more neutron rich systems favor 
larger IMF multiplicities. We select the 
classes with three IMF's (total number of fragments $N_{F}=5$) and four IMF's($N_{F}=6$),
corresponding to around 550 events 
and 250 events out of the total of 2000 events, providing so a reasonable statistics.

The charge distributions corresponding to each order in the mass hierarchy are 
shown in Fig.\ref{dis5a}, for the events with three IMF's and all entrance channel 
combinations, HH, EE and LL respectively. Let us mention that in the following, for all figures, 
the histograms brighten as the rank of IMF increases. The heaviest IMF (the rank one in hierarchy) 
can have a charge up to
$Z=16-18$ with distribution centered around $Z=6-8$ while the lightest
arrives up to $Z=8$. In the bottom row of the figure is plotted the average transverse velocity
in each charge bin calculated by considering all fragments
independent of the position in hierarchy (see Fig.\ref{dis5a} (g), (h) and (i)). 
The transverse velocity has a steep decreasing trend
with the charge, in agreement with previous findings reported in \cite{lionti2005},
and does not depend much on the asy-EOS. In fact this appears to be one feature of
the fragmentation dynamics. The larger transverse velocity of the lightest fragments
seems to indicate a reduced driving effect of the PLF, TLF ''spectators''.
All that can be related to the presence of a multifragmenting source
located in the overlap region upon which the shape instabilities of the neck
dynamics will take over.
These observations require a more detailed investigation of
the kinematic properties of fragments, once ordered in  mass. 
As we shall see the correlations between velocity and size are amplified
when analyzing the events according to the fragment rank in the hierarchy.

\begin{figure}
\begin{center}
\includegraphics*[scale=0.33]{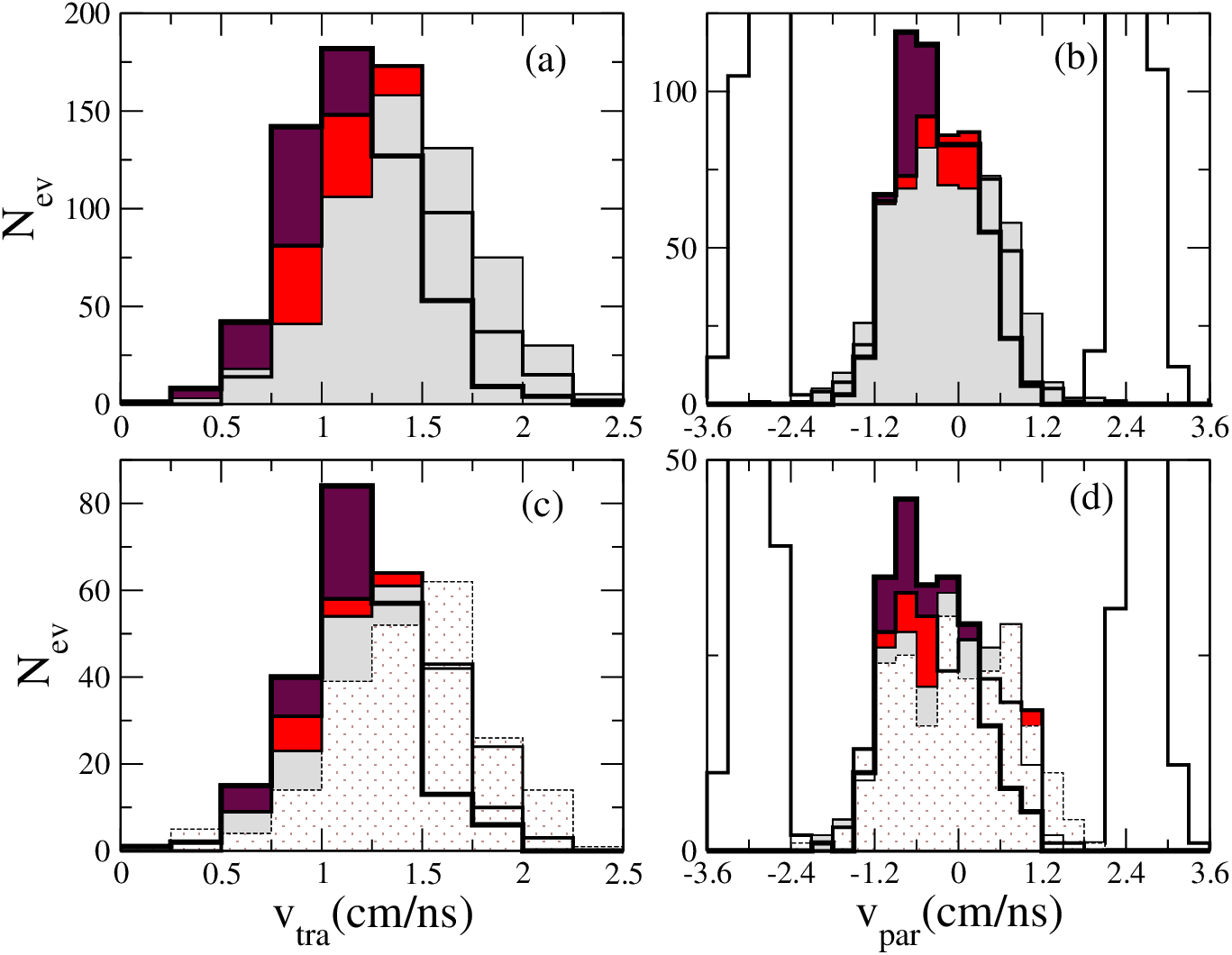}
\end{center}
\caption{HH combination and the asysoft EOS choice. Upper panels: fragmentation events
with three IMF.
(a) Transverse velocity $v_{tra}$ distributions,
(b) Parallel velocity $v_{par}$ distributions.
Bottom panels: events with four IMF.
(c) $v_{tra}$ distributions, (d) $v_{par}$ distributions.
The histograms brighten as the rank of IMF increases.
}
\label{vsoft}
\end{figure}

\begin{figure}
\begin{center}
\includegraphics*[scale=0.33]{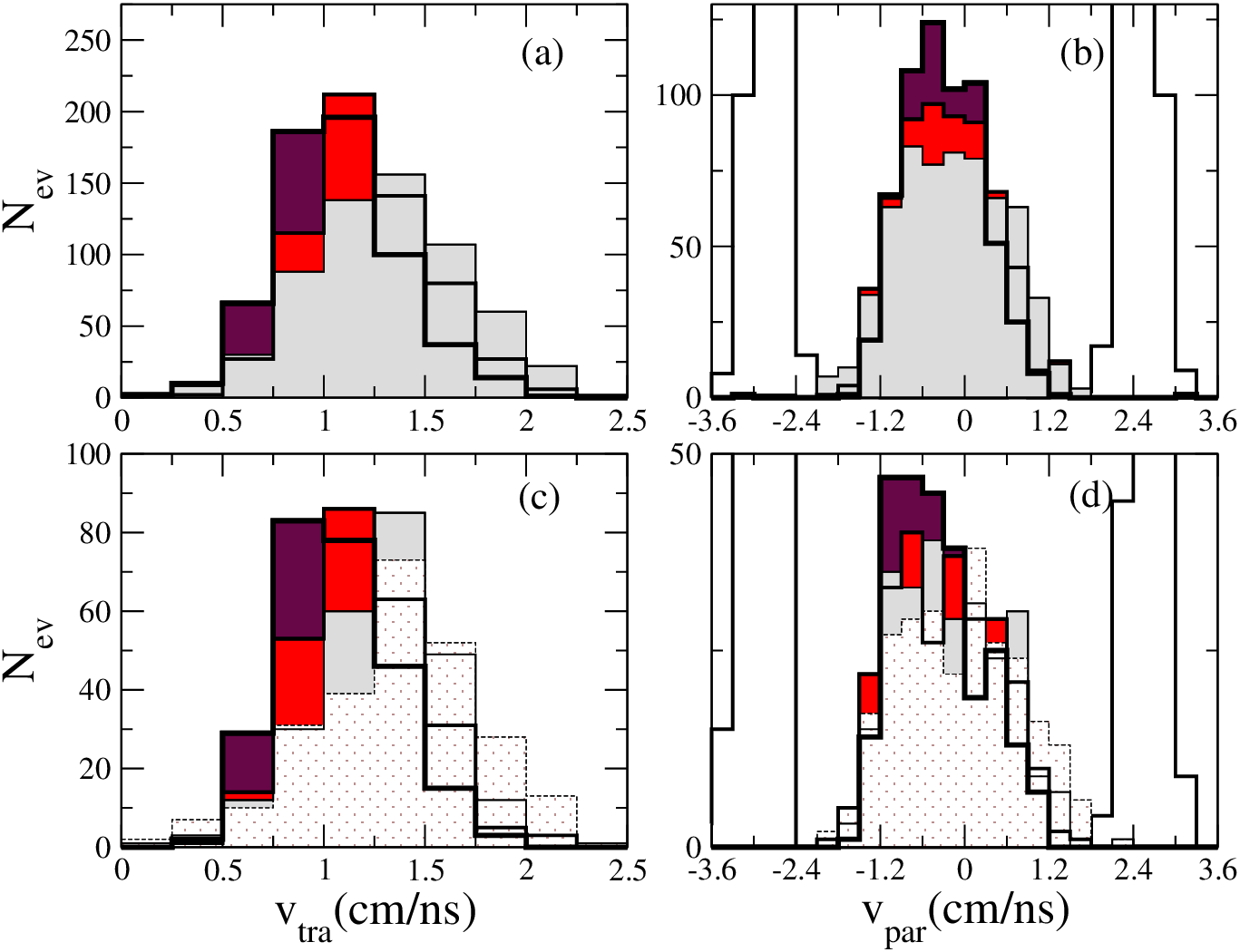}
\end{center}
\caption{Like Fig.\ref{vsoft}, for the asysuperstiff EOS choice.
}
\label{vstiff}
\end{figure}

Figs.\ref{vsoft} and \ref{vstiff} show, for asysoft and superasystiff EOS respectively,
the IMF's transverse and parallel velocity distributions 
in the case of HH combination. We also report for reference, the parallel velocity distributions 
of projectile and target like residues as they result from our calculations at these energies.

  For both classes of events considered, the transverse velocity distribution
shifts towards higher values with the IMF position in the mass hierarchy, the
lightest fragment acquiring the greatest $v_{tra}$.

This hierarchy in the velocity perpendicular to the intrinsic axis emerges as a specific
signal characterizing the transition from multifragmentation to neck fragmentation.
It can be related to the peculiar geometrical configuration of the overlapping region 
and its fast evolution. The velocity distributions along the intrinsic axis are
centered around the mid-velocity region, quite decoupled from the PLF and TLF.
This is analogous to what is observed in neck fragmentation. The parallel velocity distribution
of the lightest IMF looks broader and more symmetric around the center of
mass velocity, suggesting a dominant volume contribution of spinodal and thermal
nature to the fragment formation, like in multifragmentation.
However, it is difficult to notice any hierarchy in the IMF velocity along the intrinsic axis.
\begin{figure}
\begin{center}
\includegraphics*[scale=0.33]{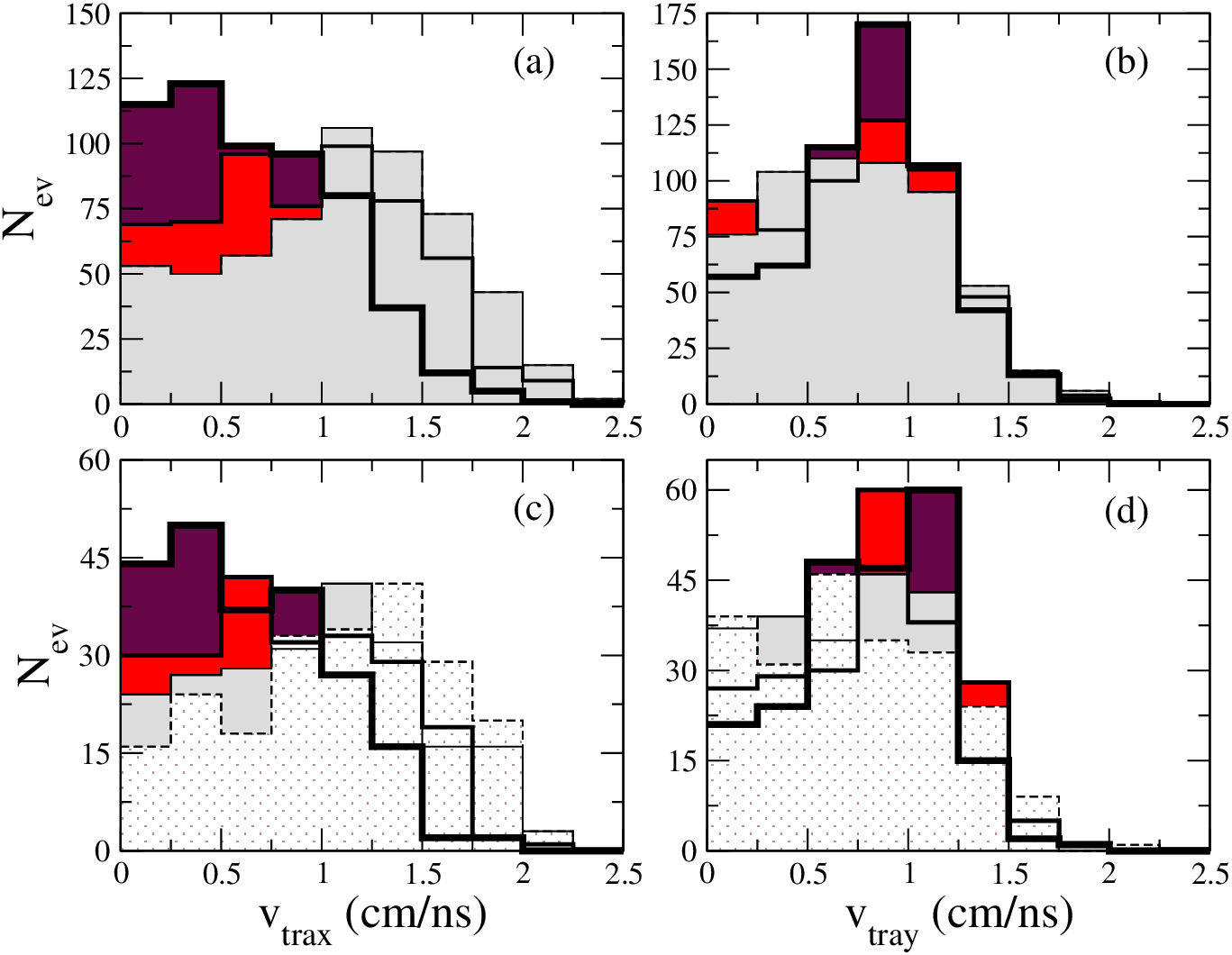}
\end{center}
\caption{The asysoft EOS and HH combination.
Transverse velocity in reaction plane distribution for fragmentation events
with three IMF (a)
and four IMF respectively (c).
Transverse velocity out of reaction plane distribution
for fragmentation events with three IMF (b) and four IMF (d).
}
\label{tra3soft}
\end{figure}

The transverse velocity features were studied in more detail, by looking separately at 
in and out of reaction plane components, $v_{trax}$ and $v_{tray}$, see Fig. \ref{tra3soft}. 
For both quantities we notice an interesting correlation
to the fragment position in the hierarchy. This signal appears rather robust, since 
in order to increase the statistics we integrate over the whole $v_{par}$ distribution of Fig.
\ref{vsoft}. For light fragments, the distribution of transverse velocity in reaction plane
appears much more extended than that associated to the component orthogonal
to the reaction plane. Just the opposite behavior is observed for the heavier fragments
of the hierarchy. The presence of the larger transverse velocities on the reaction
plane for the lightest masses in the hierarchy seems to point towards an incomplete dissipation
of the entrance channel collective energy flux.
This would be consistent with the formation of light fragments via a faster
multifragmentation
mechanism. Conversely, larger out of reaction plane velocities for the heaviest masses
reflect
their formation on longer time scales, with more initial collective energy dissipatted and 
larger Coulomb effects. In spite of this, when the two components are combined to 
generate the final distributions
in transverse velocity, the hierarchy signal is rather robust, 
see Figs. \ref{vsoft},\ref{vstiff}. This suggests that 
the rate at which the fragments
depart from the intrinsic axis is essential and this depends on their rank in mass hierarchy.

To gain more insight into the competition between thermal and dynamical, non-equilibrium effects, 
we analyzed the collective flow properties associated to the IMF's.
For each rank in mass hierarchy the transverse and elliptic flows parameters were obtained as: 
\begin{equation}
v_{1}= < \frac{p_{x}}{p_{T}} > ;\hspace{0.3cm}
v_{2}= < \frac{p_{x}^{2}-p_{y}^{2}}{p_{T}^{2}} >,
\end{equation}
where $p_{x}$ now refers to the in-reaction plane component of the momentum
perpendicular to the beam axis, while $p_{y}$ is momentum component orthogonal
to the reaction plane. Here $\displaystyle p_{T}=\sqrt{p_{x}^2+p_{y}^2}$ is the transverse momentum and the average
was performed over the number of events. The different behavior of the IMF's transverse velocity components,
for various positions in the mass hierarchy discussed above should be clearly distinguished in the flows.

\begin{figure}
\begin{center}
\includegraphics*[scale=0.33]{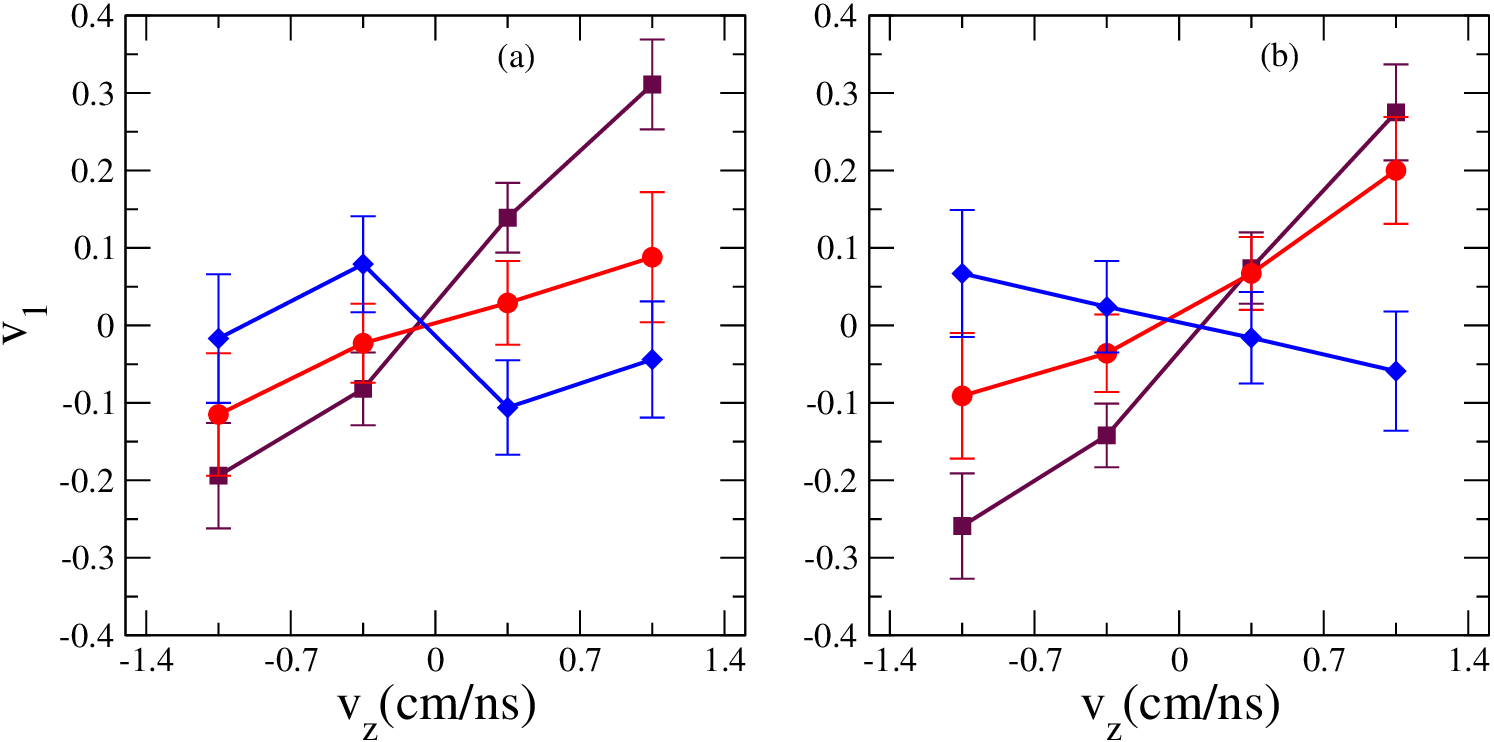}
\end{center}
\caption{Direct flow  parameter $v_{1}$ as a function of velocity along beam axis $v_{z}$.  HH combination
and events with three IMF's. Squares: heaviest IMF. Circles: second heaviest IMF. Diamonds: the lightest IMF
in hierarchy.
(a) Asysoft EOS,
(b) Asysuperstiff EOS.See the text.}
\label{v1}
\end{figure}

\begin{figure}
\begin{center}
\includegraphics*[scale=0.33]{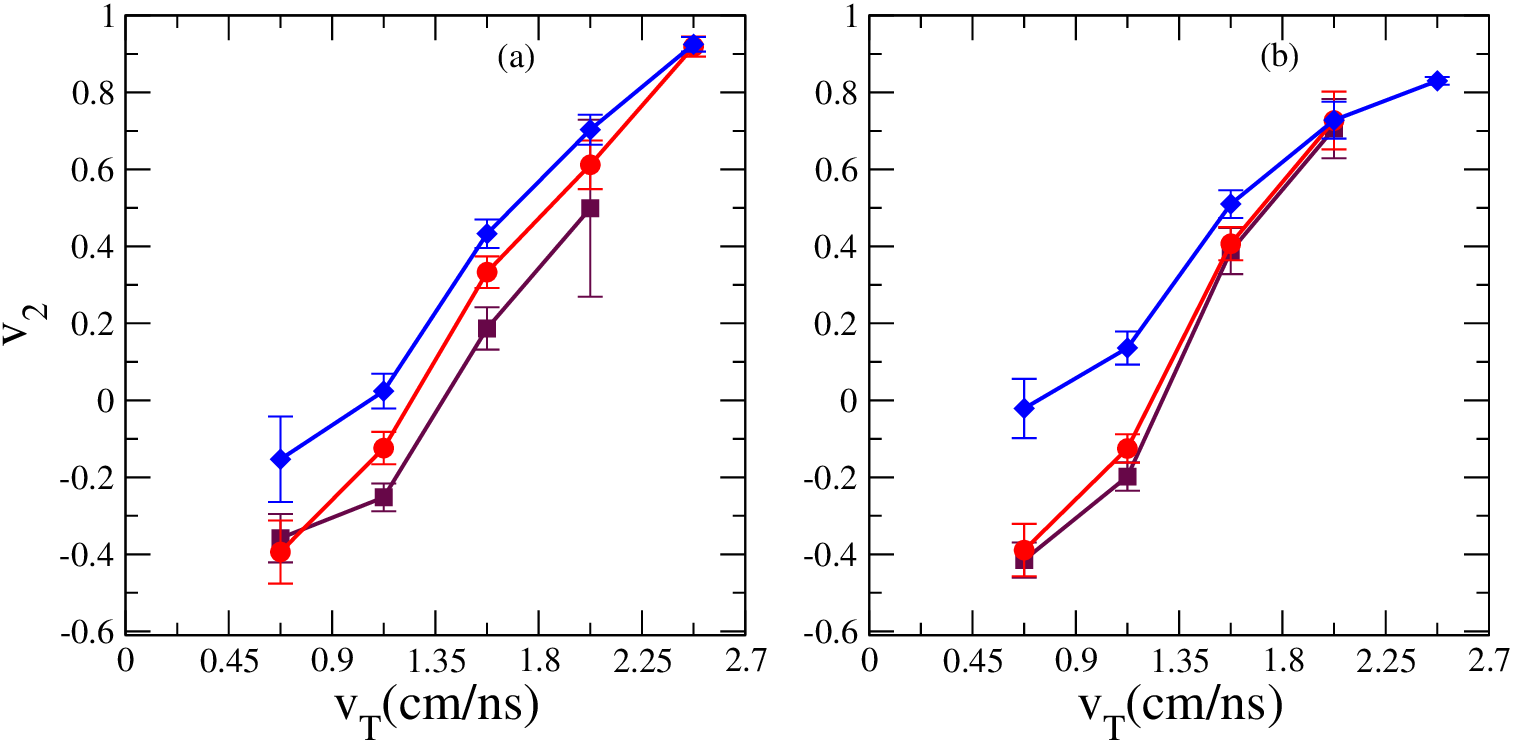}
\end{center}
\caption{ HH combination and events with three IMF's:elliptic flow anisotropy parameter $v_{2}$
as a function of transverse velocity $v_{T}$. 
Squares: heaviest IMF. Circles: second heaviest IMF.
Diamonds: the lightest IMF in hierarchy.
(a) Asysoft EOS.
(b) Asysuperstiff EOS.
}
\label{v2}
\end{figure}

In Fig.\ref{v1} we report the in plane transverse flow $v_{1}$ as a function of the fragment velocity
along beam axis $v_{z}$, for each rank in mass hierarchy with three IMF's. 
The light fragments present an almost flat, 
close to zero transverse flow, fully consistent with an early formation and decoupling 
in the mid-rapidity zone. At variance, 
the heavy fragments nicely follow positive/negative value due to the correlation to the PLF/TLF spectators.
In Fig.\ref{v2} we represent the elliptic flow parameter $v_{2}$ dependence on
the total transverse velocity 
$\displaystyle v_{T}=\sqrt{v_{x}^2+v_{y}^2}$, again for each of the three orders in hierarchy. 
For small $v_{T}$ the elliptic flow is negative, indicating a behavior dominated
by the thermal expansion out of reaction plane. 
However, the value of the anisotropy parameter $v_{2}$ increases with the position in hierarchy
in a given bin of $v_{T}$.It also rises with the transverse velocity, 
becoming positive above $v_{T} \approx 1.4 cm/ns$ . 
This feature can be related to the incomplete dissipation
of entrance channel collective energy, driving lighter fragments easier parallel
to the reaction plane. We have to look at this figure also in connection
to Fig. \ref{dis5a} (bottom panel) where the $v_{tra}$ distribution versus charge is presented.
Light fragments, more abundant at high transverse velocities clearly show positive elliptic flow
fully consistent with the analysis of Fig. \ref{tra3soft}. Heavier fragments, more abundant at lower
$v_{tra}$ nicely show more negative $v_{2}$ values, again in agreement with the results
of Fig.\ \ref{tra3soft}. All these features should be considered as specific to this fragmentation
mechanism. We stress that it is likely to exist some other production
mechanisms, with properties differing from those evidenced above, but not described by our transport model. 
These include breakup or fission of strongly deformed quasiprojectile/quasitargets, 
which take place on longer time scales as well as statistical decay of the primary fragments.
However it is hoped that a proper selection of kinematic characteristics can single out 
various classes of events. 

\section{Isospin effects}

\begin{figure}
\begin{center}
\includegraphics*[scale=0.33]{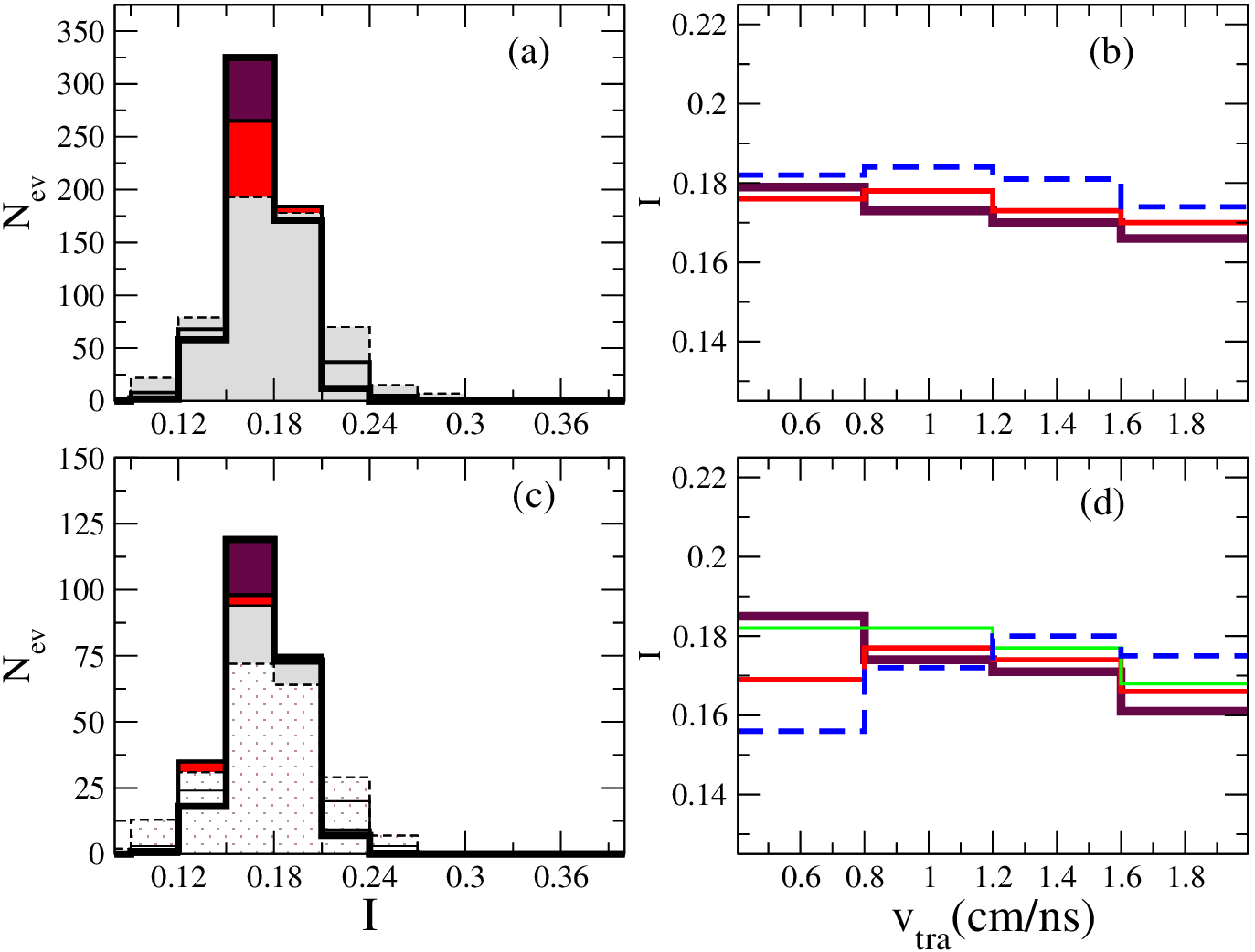}
\end{center}
\caption{HH combination in the asysoft EOS choice. Upper panels: events with three IMF.
(a) Isospin distribution.(b) Fragment isospin content as a function of transverse velocity.
Bottom panels:events with four IMF.
(c) Isospin distributions;(d) Fragment isospin content vs transverse velocity.
All histograms are like in the caption of Fig.\ref{vsoft}. The dashed lines correspond to the
lightest IMF. A thinner continuous line is associated with a lighter IMF.  
}
\label{iso1}
\end{figure}

\begin{figure}
\begin{center}
\includegraphics*[scale=0.33]{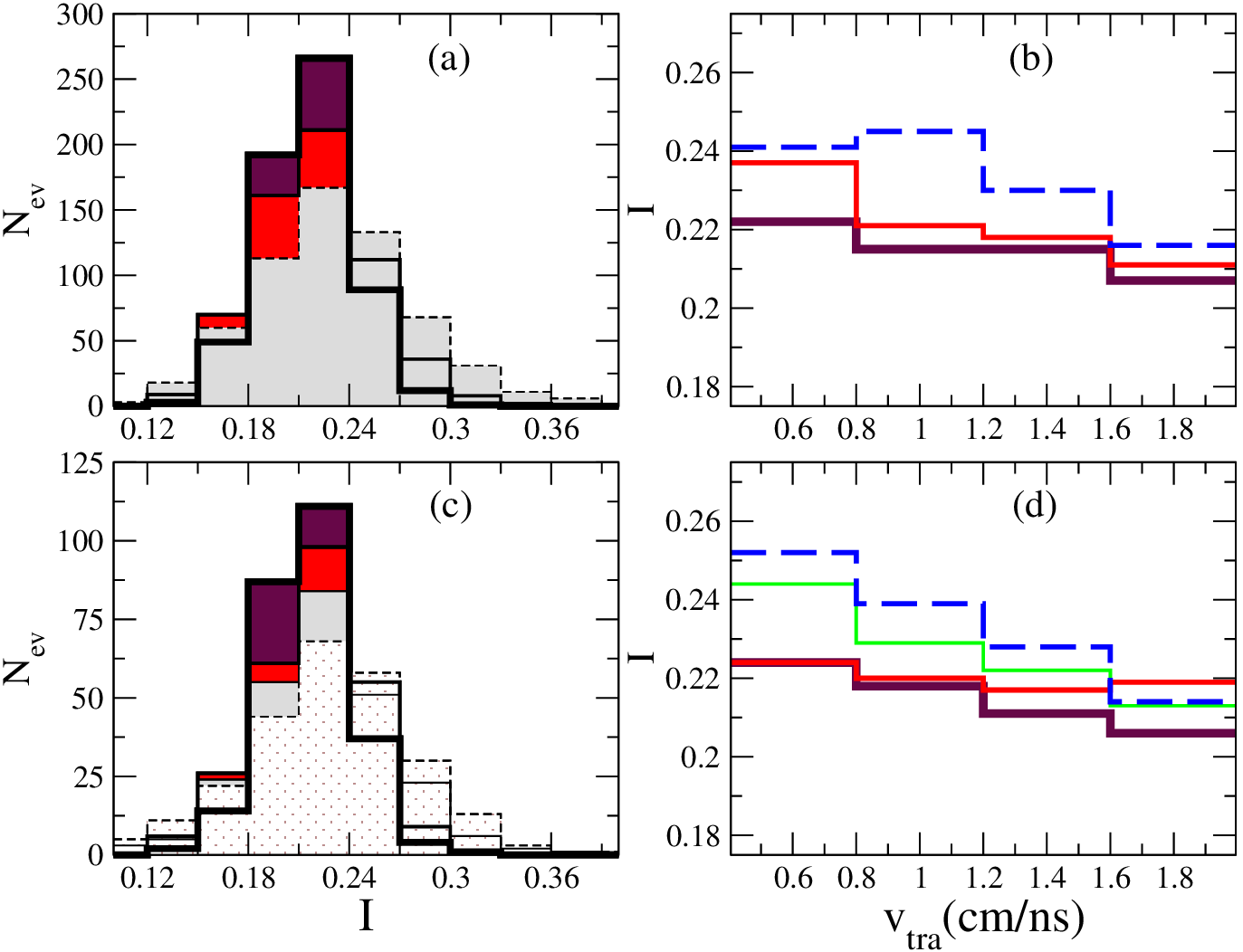}
\end{center}
\caption{
Like in Fig.\ref{iso1}, for the asysuperstiff EOS choice.
}
\label{iso3}
\end{figure}

As already noted, the features discussed above are determined mainly by the isoscalar
part of the equation of state. On top of that the symmetry energy induces various
changes on the properties related to the isospin content of the fragments. We have
extended our investigations to isospin observables studying their dependence
on the IMF position in hierarchy as well the correlation to transverse velocity.
In Figs. \ref{iso1}, (for asysoft EOS) and \ref{iso3} (for asysuperstiff EOS)
we report the asymmetry $I=(N-Z)/(N+Z)$ distribution of each IMF of the hierarchy.
The results refers again to HH system whose initial asymmetry is $I=0.194$.
Several differences between the two asy-EOS are evidenced. 
For asysoft EOS the isospin distributions are centered at a lower value
and their widths are rather narrow. At variance, for asysuperstiff EOS the 
centroids of the distributions are closer to the initial value of 
the composite system and their broader widths depend on the position in the 
mass hierarchy. For both asy-EOS the lightest IMF's
are more likely to acquire higher values of the asymmetry.
We also notice that similar results were obtained for the other entrance 
channel combinations, LL and EE respectively.

We relate these features to the differences between the two asy-EOS at 
sub-saturation densities. Clearly, a larger value of the symmetry energy will fasten 
the isospin distillation process and all IMF's reach lower and closer 
values of the asymmetry. This is the case for the asysoft EOS.
On the other hand larger values of fragment asymmetry in the case of asysuperstiff EOS
shows that this was not very effective during the formation phase. 

 The fragments continue to grow in quite low density, more asymmetric regions, as a result
of isospin migration. The differences inside the hierarchy for the latter asy-EOS
point towards different formation time scales, with the
lightest IMF finding more neutron rich environment and a distillation process not fast enough
to produce the same asymmetry for all IMF's in the event. 
However, it is interesting to remark that these fragments also acquire the 
largest transverse velocity, as it was discussed before. 
Therefore a possible scenario is that they escape faster from the active region
keeping a partial memory of the early conditions of the fragmentation. At variance,
if they have lower transverse velocity and appearing in a richer neutron region, will carry
higher asymmetry. We represent the average asymmetry as a function of transverse velocity in 
Figs. \ref{iso1} and \ref{iso3}, for asysoft and asysuperstiff EOS choice respectively. 

A decreasing trend is generally observed for the IMF's, more pronounced for asysuperstiff EOS.
Moreover, in this case, the trend is particularly evident for the lightest IMF's, 
in agreement with the previous discussion.

 For a given transverse velocity bin the asymmetry always increases with the rank
in hierarchy in the case of asysuperstiff EOS. On the other hand, for 
the asysoft EOS we cannot appreciate much such differences, all fragments reaching
almost the same asymmetry. 

 The same type of analysis has been carried out for the LL combination aiming to construct
isotopic double ratios and to study their dependence on the transverse velocities.
Concerning the fragment isotopic content, similar differences between the two asy-EOS,
as observed for the HH combination, were evidenced, in spite of the fact that Coulomb 
effects are  now more important. We also observed an analogous trend of the 
fragment asymmetry with the transverse velocity. Therefore, in this case the double ratios 
do not show appreciable differences between the two asy-EOS. The same conclusion was 
reached in central collisions \cite{colonna2008}. However, as we noticed before,
differences can be evidenced even within the same system, by comparisons between  
fragments belonging to different ranks in hierarchy for various kinematic
selections. 

\section{Conclusions and perspectives}

 In this paper, by employing a microscopic transport model, 
we unveiled new features of nuclear fragmentation in semi-central to 
semi-peripheral collisions from the study of several kinematic correlations
of intermediate mass fragments.

 At Fermi energies, an almost continuous transition with the centrality,
from multifragmentation to neck fragmentation mechanisms, is revealed. 
Good observable tracers appear to be related to the correlations between
the fragment masses,transverse velocities and isospin contents. In fact, specific 
hierarchy phenomena are signaled: the distributions of the velocity perpendicular
to the intrinsic axis of the event depend on the rank in a mass hierarchy of the event.
In the reaction plane the lightest fragments acquire greater transverse velocities,
a phenomenon observed for several mass entrance channels. This feature can be
used as an identification of the fragmentation mechanism discussed in this paper.

 Another important finding is that the fragment isospin content is 
sensitive to the position in this hierarchy and this can be related to the 
density dependence of symmetry energy at sub-saturation densities as well
as to the relative time scales for fragment formation and isospin transport.

 These observations open new opportunities from the experimental point of view.
An analysis of isospin dependent observables in correlation to position
in mass hierarchy or kinematic observables may add other constraints upon the 
behavior of symmetry energy below normal density and can provide a supplementary
support for the assumption that the IMF's form in the low density regions
of heated nuclear matter. We mention that recent experimental results, reported by CHIMERA
collaboration for the system $Sn+Ni$ at a lower energy ($35 AMeV$) \cite{defilippo10},
sustain the existence of the hierarchy in transverse velocity, as discussed in this paper.
Their analysis also signaled differences in the isospin content of IMF's when
ordered in a mass hierarchy, the lightest fragments being more asymmetric. 
This kind of observations support an asystiff-like
behavior of the symmetry energy at sub-saturation densities.

\subsection*{Acknowledgments}

The authors are grateful to E. De Filippo,\ A. Pagano,\ P.Russotto, 
and CHIMERA Collaboration for stimulating
discussions. One of authors, V. Baran thanks for warm hospitality at Laboratori
Nazionali del Sud. This work was supported in part by the Romanian
Ministery for Education and Research under the CNCSIS contract PN II ID-1038/2008.
For R. Zus this work was supported by the strategic grant POSDRU/89/1.5/S/58852, Project 
''Postdoctoral programme for training scientific researchers'' co-financed by the European Social
Found within the Sectorial Operational Program Human resources Development 2007-2013. 
\vspace{0.5cm}



\end{document}